# Natural charge spatial separation and quantum confinement of ZnO/GaN core/shell nanowires

Zhenhai Wang[1], Yingcai Fan[1], Mingwen Zhao[1,*]

*School of Physics and State Key Laboratory of Crystal Materials Shandong University, Jinan 250100, Shandong, P.R. China*

*zmw@sdu.edu.cn

We performed density-functional calculations to investigate the electronic structure of ZnO/GaN core/shell heterostructured nanowires (NWs) orientating along <0001> direction. The build-in electric filed arising from the charge redistribution at the $\{1\bar{1}00\}$ interfaces and the band offsets were revealed. ZnO-core/GaN-shell NWs rather than GaN-core/ZnO-shell ones were predicted to exhibit natural charge spatial separation behaviors, which are understandable in terms of an effective mass model. The effects of quantum confinement on the band gaps and band offsets were also discussed.

Quasi-one-dimensional semiconductor nanostructures are increasingly used in photonic and electronic devices, such as integrated nanodevices,[1] novel field effect transistors (FET)[2,3] and photovoltaic solar cells.[4,5] Recently, radially-modulated (core/shell) heterostructured nanowires (NWs) have been explored extensively as photovoltaic materials to produce high-efficiency, robust, integrated PV power sources at low cost.[4,5] Thanks to the unique structures that the interface extends along the longer axial direction of the NWs and that carrier separation takes place in the shorter radial direction, photogenerated carriers can reach the interface with high efficiency without substantial bulk recombination, which improves carrier collection and overall efficiency with respect to comparable axially-modulated heterostructures. Indeed, high electron mobility devices have been achieved in InAs/InP[6] and GaN/AlN/AlGaN[7] core/shell heterostructured NWs. First-principles studies of ZnO/ZnS,[8,9] Si/Ge[10,11] and GaN/GaP[12] core/shell NWs have demonstrated the accumulation of ZnO-core, Ge-core hole gas and GaN-core electron gas, in agreement with experimental observations. Moreover, the band offset of core-shell NWs can be modulated by the quantum confinement effect and the strain induced by the lattice mismatch between the two components, yielding a board range of band gaps, which facilitates light adsorption of PV devices.

The core-shell NWs with type-II band alignment are advantageous in promising solar energy applications such as water splitting, dye-sensitized and even regular solar cells.[13-15] The synthesis of ZnO/GaN core/shell NWs and nanotubes with type-II band alignment have been reported.[16-18] The spectrum of GaN-core/ZnO-shell nanotubes exhibit band energies from 1.9 eV to 3.6 eV, which are in the visible region.[17,18] Theoretical study of GaN/ZnO super-lattice and random-alloy has provided important insights and guidelines for designing band gap reduction.[19] However, the band offset and build-in electric filed at the interface of ZnO/GaN core/shell NWs have never been reported. Compared with other core/shell NWs,[8,9] ZnO/GaN has a very small lattice mismatch (~1.86%) and thus low interface strain. The strain effect on the electronic structure are very limited, offering an ideal model to reveal the radial quantum-confinement effects on the band offset, built-in electric field, and electron-hole separation of core/shell NWs.

In this contribution, we performed first-principles calculations to investigate the electronic structure of heterostructured ZnO-core/GaN-shell and GaN-core/ZnO-shell NWs. We found that core/shell NWs exhibit a band offset larger than that of bulk superlattice due the quantum confinement effects.

The ZnO-core/GaN-shell NWs have natural charge spatial separation with holes and elections being confined in core and shell regions. The build-in electric filed induced by the charge redistribution in the regions near the interfaces facilitates the hole-electron separation, making the NWs promising candidate nanomaterials for PV applications.

Our first-principles calculations were performed using the projected augmented wave (PAW) basis combined with the generalized gradient approximation (GGA) in the form of Perdew Burke Ernzerhof (PBE) for the electron exchange-correlation functional, which were implemented in the Vienna ab initio simulation package(VASP).[20-22] A GGA+U scheme was employed to improve the electric structure calculations. The U values used in this work are 4.7 eV and 3.9 eV for the *d* orbitals of Zn and Ga atoms, respectively.[23] The band gap values of w-ZnO and w-GaN obtained from the present GGA+U calculations are 1.42 eV and 2.47 eV, respectively, in contrast to the experimental values, 3.25 eV and 3.19 eV. More accurate schemes, such as quasiparticle GW approximation or HSE hybrid functional are needed to get band gap values more comparable to experimental data.

We calculated the electrostatic potential $V(\vec{r})$ by solving Poisson equation. The planar-averaged potential $\bar{V}(z)$ was then obtained using the expression:

$$\bar{V}(z) = \frac{1}{S}\int_S V(\vec{r})dxdy \quad \ldots (1)$$

where $S$ represents the area of a unit cell in the plane parallel to the interface (xy-plane). The macroscopic average $\bar{\bar{V}}(z)$ is accomplished by averaging $\bar{V}(z)$ at each point over a distance corresponding to one period (L):

$$\bar{\bar{V}}(z) = \frac{1}{L}\int_{-L/2}^{L/2} \bar{V}(z')dz' \quad \ldots (2)$$

The valence band offset ($\Delta E_V$) of ZnO/GaN heterostructures can be evaluated using the equation:

$$\Delta E_{VBO} = \Delta\bar{\bar{V}}|_{ZnO/GaN} + (E_{VBM} - \bar{\bar{V}})|_{ZnO} - (E_{VBM} - \bar{\bar{V}})|_{GaN} \quad \ldots (3)$$

where the first term represents the difference between the $\bar{\bar{V}}(z)$ values of the two components in the heterostructures, while the second and third terms are the difference between VBM energy ($E_{VBM}$) and $\bar{\bar{V}}(z)$ of the corresponding isolated components in bulk crystal.

For propose of comparison, we first calculated a ZnO/GaN superlattice with nonpolar ($1\bar{1}00$) interface structure, as shown in FIG.1(a). The ($1\bar{1}00$) interface structure is energetically more favorable than the polar $(0001)/(000\bar{1})$ interface.[24] In our calculations, the supercell has 8 ZnO bilayers and 8 GaN bilayers, and the stable configuration contains minimum number of "wrong bonds" (i.e. Ga-O and Zn-N bonds) at the interface. The bond lengths of the "wrong bonds" are very close to those of Zn-O and Ga-N (see Table 1), suggesting that the interface is less strained.

The electrostatic potential profiles along the growth direction are shown in FIG. 2(a). Obviously, built-in field was set up across the interface with the strength of 0.042 V/Å. This is related the charge redistribution in the region near the interface. The electronegativity difference between N and Zn atoms is 1.39 (Pauling scale), smaller than that between O and Ga atoms 1.63. Therefore, there is a net electron transfer from GaN segment to ZnO segment, inducing electric filed pointing to GaN segment. The charge density distribution shows that the valence band minimum (VBM) is mainly contributed by the N $2p$ orbitals and the conduction band maximum (CBM) arises from the O $2s$ and Zn $3s$ orbitals, implying type-II band alignment features. The $\Delta E_{VBO}$ calculated using Eq.(3) is 1.04eV with higher GaN VBM. It is noteworthy that the band offset values depend on the interface structures and chemical components. Our value is in accordance with the results of some experimental[25] and theoretical works[19], which are 0.7-1.3eV.

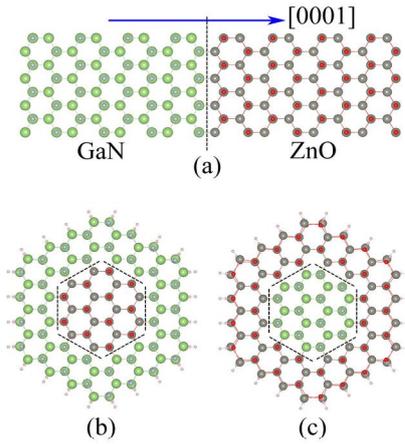

FIG. 1 (Color online) Geometric structures of (a) GaN:ZnO ($1\bar{1}00$) interface along [0001] growth direction; (b)ZnO-core and (c) GaN-core NWs with hydrogenated facets. Pink, red, green, blue and gray spheres represent H, O, N, Zn and Ga atoms, respectively. The dashed lines mark out the {$1\bar{1}00$} interfaces.

Table I. Bond lengths of super-lattice (SL) interface structure and Core-shell nanowires in different section.

|  | $d_{Zn-O}$(Å) | $d_{Ga-N}$(Å) | $d_{Zn-N}$ (Å) | $d_{Ga-O}$ (Å) |
|---|---|---|---|---|
| SL | 1.993[b], 2.023[i] | 1.972[b],1.946[i] | 1.980 | 1.946 |
| ZnO-core | 1.993[b],2.019[i] | 1.975[b],1.966[i], 1.981[s] | 1.981 | 1.946 |
| GaN-core | 1.966[b],2.021[i], 2.101[s] | 1.969[b],1.948[i] | 1.984 | 1.955 |

[b]bulk    [i]interface    [s]surface

We then modeled the morphologies of ZnO/GaN core-shell heterostructured NWs using hexagonal prisms orientating along <0001> direction enclosed by six {$1\bar{1}00$} facets, as shown in FIG.1(b) and (c). All the surface dangling bonds are saturated by hydrogen atoms. Both ZnO-core/GaN-shell and GaN-core/ZnO-shell heterostructured NWs with six nonpolar {$1\bar{1}00$} interfaces (denoted as ZnO-core and GaN-core NWs hereafter) were considered. Due to computational limitations, the thicknesses of core and shell region contain only two bilayers, respectively, with the NW diameter of about 2.4nm. Similar to the cases of ZnO/GaN superlattice, slight distortion occurs near the interfaces with the deviation less than 1.5% for both NWs (see Table I). The electrostatic potential profiles along the radial direction are plotted in FIG. 2(c). Obviously, built-in filed is formed across the interfaces, pointing outward (or inward) for ZnO-core (or GaN core) heterostructured NWs, respectively. The $\Delta E_{VBO}$ evaluated using Eq.(3) is 1.44eV for ZnO-core NW, higher than that of ZnO/GaN superlattice 1.04eV. This difference can be attributed to the size-confinement effect of the NW which modulates the electron density distribution and thus the electrostatic potential.

The band structures of ZnO-core and GaN-core NWs are plotted in FIG.2 (d). Both NWs have direct band gaps at the Γ point. The conduction bands near the Fermi level are dispersed throughout the Brillouin zone (BZ) and exhibit near free-electron features, which facilities the motion of electrons excited to these bands. The effective masses of lowest conduction bands are $0.505m_e$ and $0.497m_e$ for ZnO-core and GaN-core NWs, respectively. Compared with conduction bands, the valence bands near the Fermi level are rather flat. Partial electron density of states (PDOS) projected onto different atoms indicate that the conduction bands near the Fermi level arises mainly from the $3s$ orbitals of Zn and O atoms for ZnO-core NW, but for GaN-core NW, the $3s$ orbitals of the four kinds of atoms contribute to the conduction bands. The valence bands near the Fermi level come from the N $2p$ orbitals for both NWs. Therefore, the extended $3s$ orbitals and localized N $2p$ orbitals are responsible for the different profiles between the conduction and valence bands. Moreover, the PDOS of ZnO-core NWs exhibits a clear feature of charge spatial separation with electrons and holes being confined within core (ZnO) and shell (GaN) regions, respectively.

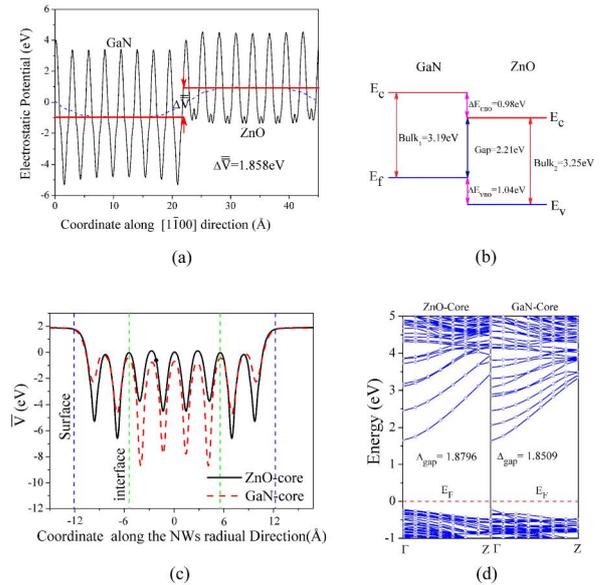

FIG. 2 The electrostatic potential and band structures of super-lattice and core-shell NWs: (a) The electrostatic potential across the GaN/ZnO ($1\bar{1}00$) interface; (b) A schematic type-II band alignment of ZnO/GaN super-lattice; (c) The electrostatic potential along a radius direction of core-shell NWs; (d) Band structures of ZnO-core and GaN-core NWs.

To visualize the electron-hole separation features, we plotted the spatial distribution of charge density distributions for the bands near the Fermi level in FIG.3(a) and (b). It is clear that for ZnO-core NW the valence bands (VBM and VBM-1) have electron wave functions being confined within the shell (GaN) region; whereas the wave functions of the conduction bands (CBM, CBM+1, and CBM+2) mainly locate in the core (ZnO) region. For the GaN-core NW, however, the CBM wave function extends throughout the core and shell regions which breaks the spatial separation between valence bands and conduction bands near the Fermi level.

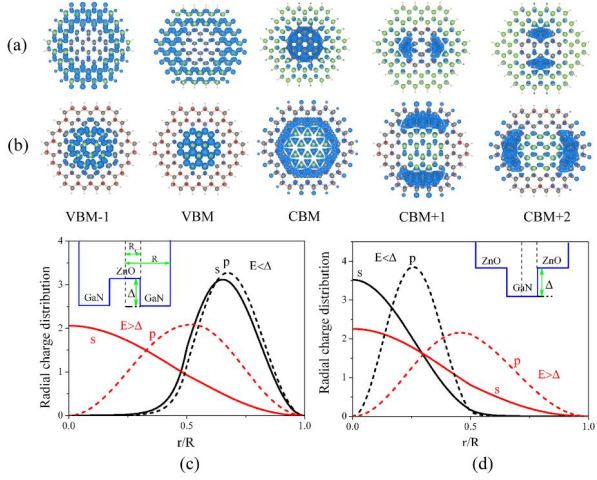

FIG. 3 (a),(b): the charge distributions for the states near the Fermi level for different Core-Shell NWs. (c),(d): the radial charge distribution for the *s* and *p* states in the effective mass model for $R_c/R=0.5$, $E=\Delta/5$ (black curves) and $E=5\Delta$ (red curves), respectively. $\Delta$, $R_c$ and $R$ are the potential barrier, the core radius and the radius of the NWs, respectively.

This is understandable in terms of a simple effective mass model of one electron in a confining potential with cylindrical symmetry V(r).[12] A potential well (for GaN-core) or barrier (for ZnO-core) is introduced into the center of the cylindrical potential to represent different core/shell heterostructures. Assuming that the wave functions are in the form of $\Psi(r,\theta,z)= \varphi(r)e^{i(l\theta+kz)}$, the one-electron Schrödinger equation can be separated into the free motion along the axial (z-) direction and the radial equation:

$$\frac{\partial^2 \varphi}{\partial r^2}+\frac{1}{r}\frac{\partial \varphi}{\partial r}+(\frac{2m^*}{\hbar^2}(E-V(r))-\frac{l^2}{r^2})\varphi=0 \quad (4)$$

where $m^*$ is the effective mass and $l$ ($l =0, \pm1, \pm2 \ldots$ ) is the angular momentum. The Eq.(4) can be solved using Bessel functions. For the central well case, we have the wave function $\varphi(r) = cJ_l(\lambda r)$ in the core region and $\varphi(r) = c_1J_l(kr) +c_2Y_l(kr)$ in the shell region, where $k = (2m^*(E-\Delta))^{1/2}/\hbar$ and $\lambda = (2m^*E)^{1/2}/\hbar$ are the length scales of Bessel functions. This model describes a set of parabolic sub-bands with corresponding eigenvalues $E_{nl}$ at the BZ center. $E_{nl}$ can be obtained by using the boundary conditions such that the wave function and its derivative are continuous at the interface ($R_c$) and vanish at the edge (*R*). FIG.3(c) and (d) shows the radial charge distribution for the confined (E<$\Delta$) and unconfined (E>$\Delta$) states with $l$ =0 and 1, which are referred to as (*s*, *p*) and (*s′*, *p′*), respectively. There is a clear spatial separation between confined (*s*, *p*) states and unconfined (*s′*, *p′*) states for the central barrier case. However, for the central well case, the charge density of the unconfined *s′* state and confined *s*, *p* states locate in the core region. This is in good agreement with the results of first-principles calculations. Additionally, according the effective mass model, the *p* states are two-fold degenerated ($l$ =±1). Our first-principles calculations revealed the same features that the CBM+1 are energetically degenerated for both ZnO-core and GaN-core cases.

In summary, our first-principles calculations show that the <0001>-orientated ZnO-core/GaN-shell heterostructured NWs have natural charge spatial separation with electrons and holes being confined within core and shell regions, respectively. The built-in electric field pointing toward shell region formed across the interface facilities the electron-hole separation. The electrons excited into the conduction bands near the Fermi level exhibit near-free-electron features with effective mass of about 0.5 $m_e$. The band offset and band gap of these core-shell heterostructured NWs can be modulated by the quantum confinement effects, resulting in broad absorption band ranging from 2.21eV to 3.71eV. These results suggest that the ZnO-core/GaN-shell NWs are promising candidates for new generation solar cells.

## ACKNOWLEDGEMENTS

This work is supported by the National Natural Science Foundation of China (No. 10974119), the National Basic Research 973 Program of China (No. 2005CB623602), the Independent Innovation Foundation of Shandong University (IIFSDU, No. 2009JQ003), and the Graduate Independent Innovation Foundation of Shandong University (GIIFSDU).